\documentclass[preprint]{aastex}

\begin{document}

\title{Do mirror planets exist in our solar system?}

\author{R. Foot$^{1,3}$ and Z. K. Silagadze$^{2,4}$}
\affil{ $^1$School of Physics, Research center for High Energy Physics, \\
University of Melbourne, Victoria  3010 Australia \\
$^2$Budker Institute of Nuclear Physics, 630 090 Novosibirsk, Russia }
\email{$^3$Foot@physics.unimelb.edu.au, $^4$Silagadze@inp.nsk.su}

\date{}








\begin{abstract}
Mirror matter is predicted to exist if parity is an unbroken
symmetry of nature. Currently, there is a large amount
of evidence that mirror matter actually exists coming
from astrophysics and particle physics.
One of the most fascinating (but speculative) possibilities
is that there is a significant abundance of mirror matter
within our solar system. If the mirror matter
condensed to form a large body of planatary or stellar
mass then
there could be interesting observable effects.
Indeed studies of long period comets
suggest the existence of a solar companion
which has escaped direct detection and is therefore
a candidate for a mirror body.
Nemesis, hypothetical ``death star'' companion of
the Sun, proposed to explain biological mass extinctions,  
may potentially be a mirror star.
We examine the prospects for detecting these objects if they do
indeed exist and are made of mirror matter.

\end{abstract}


One of the most interesting candidates for  
dark matter coming from particle physics is ``mirror matter''.
Mirror matter is predicted to exist if
parity is a symmetry of Nature (Lee and Yang 1956, Kobzarev et al. 1966,
Pav\v{s}i\v{c} 1974,  Foot et al. 1991). The idea is 
that for each ordinary particle, such as the photon, electron, proton
and neutron, there is a corresponding mirror particle, of 
exactly the same mass as the ordinary particle.
The fundamental interactions of the mirror particles precisely
mirrors those of the ordinary particles.
For example, the mirror proton interacts with the mirror photon
in precisely the same way in which an ordinary proton interacts
with an ordinary photon.
The mirror particles are not produced
in laboratory experiments just because they couple very
weakly to the ordinary particles. 
In the modern language of gauge
theories, the mirror particles are all singlets under 
the standard $G \equiv SU(3)\otimes SU(2)_L \otimes U(1)_Y$
gauge interactions. Instead the mirror
particles interact with a set of mirror gauge particles,
so that the gauge symmetry of the theory is doubled,
i.e. $G \otimes G$ (the ordinary particles are, of 
course, singlets under the mirror gauge symmetry) (Foot et al. 1991).
Parity is conserved because the mirror particles experience
$V+A$ mirror weak interactions
and the ordinary particles experience the usual $V-A$ weak
interactions.  Ordinary and mirror
particles interact with each other predominately by
gravity only. 

While mirror matter has always been extremely well
motivated theoretically, it is only in relatively recent times
that the experimental and observational evidence for it
has accumulated to the point where an impressive case
for its existence can be made (for a review of the current
status of mirror matter, see Foot 2001b). 
First, it provides a natural candidate for dark matter. Mirror matter
is naturally dark and stable and appears to have the
necessary properties to explain the dark matter inferred
to exist in the Universe (Blinnikov and Khlopov 1982, 1983, 
Kolb et al. 1985, Khlopov et al. 1991, Hodges 1993, Matsas et al. 1998,
Bell and Volkas 1999, Berezinsky and Vilenkin 2000, 
Berezhiani et al. 2000).
On galactic scales, there is evidence from a recent weak
microlensing study (Erben et al. 2000, Gray et al. 2001) for large 
clumps of invisible matter which might be a mirror 
galaxy (or galaxy cluster) (Foot 2001b).
Within galaxies such as our own
Milky way, mirror matter may be the dominant component of the halo,
thereby explaining the MACHO observations (Silagadze 1997, 
Blinnikov 1998, Foot 1999, Mohapatra and Teplitz 1999)
\footnote{The conventional red, brown 
or white dwarf interpretation of these MACHO events have
real problems (see e.g. Freese et al. 1999). It is also possible
that the MACHO events are due to lens in the LMC (and
not actually in the halo of our galaxy), however this interpretation
also is problematic (see for example, Gyuk et al. 1999).}. 
On small scales (such as solar system scale)
systems containing ordinary and mirror matter could exist but
it is likely that they should be quite unequally mixed
(e.g. 99\% ordinary matter and 1\% mirror matter). This is
because ordinary and mirror matter are naturally segregated on
small scales as they don't have common dissipative interactions
(Blinnikov and Khlopov 1982, 1983, Kolb et al. 1985, 
Khlopov et al. 1991). 
In fact, the strange properties of some of the extrasolar
planets may be explained more naturally if they are mirror
planets (Foot 1999b, 2001a).  
Furthermore, recent Hubble Space Telescope star 
count results show the deficit of local
luminous matter (Blinnikov 1999, 2000; However there is some controversy 
with Hipparcos satellite data, see Holmberg and Flynn 1998),
expected if the population of the mirror
stars in the galactic disk is numerous enough (Blinnikov and Khlopov
1982, 1983).

On quite a different tack, there is
evidence for mirror matter
coming from the solar and atmospheric neutrino anomalies 
(Foot et al. 1992, Foot 1994, Foot and Volkas 1995. For a review of the
neutrino physics anomalies, see e.g. Langacker 1999).
Ordinary and mirror neutrinos
are maximally mixed with each other if neutrinos have mass
(Foot et al. 1992, Foot 1994, Foot and Volkas 1995).
The maximal $\nu_e \to \nu'_e$ (the $'$ denote the mirror
particle) oscillations predict an  approximate 
$50\%$ $\nu_e$ flux reduction
thereby explaining the solar neutrino experiments
while the maximal $\nu_\mu \to \nu'_\mu$ oscillations
predict the up-down neutrino asymmetry observed 
in Super-Kamiokande (Fukuda et al. 1998a, 1998b)
(see e.g. Foot et al. 1998, Foot 2000,  Fornengo et al. 2000 for 
a fit of maximal $\nu_\mu \to \nu'_\mu$ oscillations to the data).
The idea is also compatible with the LSND experiment
(Foot et al. 1992, Foot 1994, Foot and Volkas 1995). 
Interestingly, maximal ordinary - mirror neutrino oscillations 
do not pose any problems for big bang nucleosynthesis (BBN)
and can even fit the inferred primordial abundances better
than the standard model (Foot and Volkas 1997, 2000).

Finally there are several other interesting effects of mirror
matter which have been discussed such as 
photon - mirror photon kinetic mixing (Holdom 1986, Glashow 1986,
Carlson and Glashow 1987, Collie and Foot 1998), Higgs 
- mirror Higgs mixing (Foot et al. 1991, H. Lew, unpublished) and
possible ordinary - mirror particle interactions (Silagadze 1999)
expected in currently popular models of large extra dimensions
(Akama 1982, Rubakov and Shaposhnikov 1983, Arkani-Hamed et al. 1998).
It should also be noted that there are variants of
the mirror matter idea where the mirror symmetry is 
assumed to be spontaneously broken (Barr et al. 1991, Akhmedov et al.
1992, Foot and Lew 1994, Berezhiani and Mohapatra 1995,
Berezhiani et al. 1996, Berezhiani 1996, Mohapatra and Sciama 1998,
Lindebaum et al. 2000).

Given the possibility that many nearby stars have ``hot jupiters'',
which may really be ``cool mirror planets'', it is possible that
there are also mirror stars/planets/comets etc gravitationally 
bound to our sun. Of course, any very nearby large planet would
have been detected via its gravitational influence. A more
distant companion is a priori a fascinating possibility. 
In fact there is some evidence for the existence of such
objects from biological mass extinctions and recent studies of
long period comets as we now discuss.

Over the past 15 years or so there has been speculation that 
there is a companion star to the sun, called ``Nemesis''
(Whitmire and Jackson 1984, Davis et al. 1984). 
The motivation for Nemesis was based
on studies suggesting that biological mass extinctions displayed
some periodicity (on a time scale
of about 26 million years) which required an extraterrestrial 
cause (Raup and Sepkoski 1984). It was also argued
that the ages of craters displayed a similar periodicity
(Rampino and Stothers 1984, Alvarez and Muller 1984).
The idea is that Nemesis would have a moderately eccentric orbit
with an orbital period of 26 million years,
which would periodically disturb the Oort cloud and cause comets to enter
into the inner solar system and trigger
the mass extinctions. 
Subsequent searches for Nemesis failed
to find it (Perlmutter 1986) and also some studies suggested that its
orbit was likely to be unstable (see e. g. Clube and Napier 1984).
However if the orbit is near the galactic plane, 
the {\it current} Nemesis's
lifetime can be as big as $10^9$ years (Hut 1984, Torbett and Smoluchowski
1984, Vandervoort and Sather 1993). 
This lifetime is not long enough for Nemesis to 
have been in such a large orbit at the 
formation of the solar system, about $5\times 10^9$ years ago. 
However at the formation of the solar system, 
at which time Nemesis was
also presumably formed, the orbit may have
been much tighter, expanding to the present orbit as a 
consequence of tidal perturbations from passing stars
and molecular clouds (Hut 1984).
It has been argued that the perturbations by
gigantic molecular clouds may be the most serious threat for stability
of Nemesis (Clube and Napier 1984), but it has also
been argued that the very diffuse nature of these massive clouds
greatly reduces the possible effect (Morris and Muller 1986).

Recently, new much more direct 
evidence for planetary or stellar companions to the sun
has also emerged. Two groups 
(Murray 1999, Matese et al. 1999) have studied the orbits of
long period comets. They find that there is 
a statistically significant excess of 
aphelion distances of long-period comets aligned on a great circle
(for comets in the 30k-50k A.U. range).
The approach of the two groups was quite different, with the Murray 1999
taking a subsample of the most accurately observed long period comets
while Matese et al. 1999 used a larger sample, but included less
well observed comets.
Apparently, the two groups find somewhat different great circles, which
can mean several things. It might mean that there are two companions,
or only one companion (if one of the groups is mistaken) or
no such companion (if they both screwed up).
For example, the study of Murray 1999 finds that
the data suggests the existence of a large planet or star 
with orbital period of around 6 million years (which
implies a distance from the sun of about 32000 A.U.
for a circular orbit).
The analysis suggests that the orbital plane of the
companion planet/star was inclined at roughly $35^o$
to the galactic plane with a retrograde orbit. 
Interestingly, both of these characteristics, the relatively low
inclination to the galactic plane and the retrograde
orbit were already identified as necessary conditions
for the stability of such orbit
(Hut 1984, Torbett and Smoluchowski 1984, Vandervoort and Sather 1993).
Thus, it seems to be possible that the
hypothetical planet/star identified in Murray 1999
was an original member of the solar system.
Clearly, further data should clarify whether
such companions really exist.


If companion stars/planets do exist, then it is possible 
that they are
light enough to be below the hydrogen burning threshold
and may have escaped
detection. However, another possibility is that the 
companion objects may be made of mirror matter (the
possibility that Nemesis exists and is made of 
mirror matter was earlier discussed in Silagadze 2001
\footnote{
The possibility that the protosolar nebula could
contain ``shadow'' matter and its evolution could lead to the formation
of some mirror solar objects, like Nemesis, was also mentioned in 
Kolb et al. 1985. But this idea was not further developed in Kolb 
et al. 1985 and even taken seriously, because it was thought that 
big bang nucleosynthesis data excludes the ``shadow world'' with 
completely symmetric microphysics.}).
This will give a simple 
explanation  for why their orbital plane is inclined with respect to 
the ecliptic 
(naturally, tidal perturbations may have modified their orbits somewhat
over time too).
Indeed, because ordinary and mirror matter couple 
together mainly by gravity, it is natural for
the ordinary and mirror parts of nebula (from which
the solar system was made) to have 
different initial conditions, like angular momentum.
If the galaxy contains a significant amounts of mirror matter, such 
mixed protosolar nebula can be formed, for example, during 
inter-penetration of ordinary and mirror giant 
molecular clouds (Khlopov et al. 1991). 

Of course it is certainly true that
if there is a mirror matter companion within our
solar system then its 
existence will be challenging to establish. Nevertheless it is 
important to keep in mind that this possibility, which might be true,
is in principle a testable hypothesis!

First of all let us mention some indirect checks. If the Sun-Nemesis 
constitute a mixed binary system there will be other similar star systems
around. We have already mentioned strange properties of some recently
observed extrasolar planets and their interpretation as mirror planets 
orbiting ordinary stars (Foot 1999, 2001a). One can imagine 
a reversed situation: an ordinary planet orbiting mirror star. Remarkably 
eighteen ``isolated planetary mass objects'' were actually discovered 
(Zapatero Osorio et al. 2000, Lucas and Roche 2000; See also 
Tamura et al. 1998) in $\sigma$
Orionis star cluster. Instead of being really isolated, which will 
challenge conventional theories of planet formation, these objects could
be ordinary Jovian type planets orbiting invisible mirror stars 
(Foot et al. 2000a).
This idea can be tested by searching for a periodic Doppler shift of
absorption lines in the planet emanation spectra (Foot et al. 2000a), 
or/and by Planetary Microlensing technique
(Mao and Paczy\'{n}ski 1991; For recent review see Sackett 1999).

Photon-mirror photon mixing can effect the orthopositronium lifetime 
(Glashow 1986) and lead to an interesting resolution of the orthopositronium
lifetime puzzle (Foot and Gninenko 2000). 
If the mixing parameter has indeed the magnitude
required for the mirror world interpretation of the orthopositronium anomaly
(and this will be experimentally tested in future vacuum cavity experiments),
a new window will be opened in mirror matter searches in the solar system. As
mirror meteoroids would effectively interact with Earth's atmosphere in this
case, releasing most of their kinetic energy in the atmosphere and possibly 
ending in atmospheric explosion (Foot 2001b, Foot and Gninenko 2000).
In such ``Tunguska-like''
events neither meteoroid fragments nor any significant crater would be found.
Also, any ordinary matter accreted onto the mirror companion
can potentially become hot due to the coupling of mirror
matter to ordinary matter via the photon -mirror photon
mixing. This may make the mirror companion potentially
observable (and may be appear to have the
characteristics of a strange type of white dwarf, especially 
if the companion object is of stellar weight) (Foot et al. 2000b).

Another means of investigating
the Nemesis hypothesis is provided by exploration of
cratering rates of the nearby celestial bodies such as the Moon and the Mars.
It was argued (Muller 1993)
that the age distribution of craters on the Moon
can be studied by using lunar spherulus. A pilot study had been already 
performed (Culler et al. 2000)
using 155 spherulus from the lunar soil delivered by 
Appolo-14 mission. The results are promising. From 3~Gyr ago
until about 0.4~Gyr ago
the inferred cratering rate gradually decreases. This is consistent 
with expectation that the density of potential impactors (asteroids and 
comets) should decrease as time goes by, because Jupiter slowly eliminates
them by deflecting them into the Sun or ejecting them out of the solar system.
At 0.4~Gyr, however, the rate suddenly increases by a factor of $3.7\pm 1.2$
and returns to the level it had 3~Gyr earlier. This fact has ``a ready 
explanation'' (Culler et al. 2000) 
in the framework of the Nemesis hypothesis. One can
imagine that just about 0.4~Gyr ago the Nemesis was perturbed into a more
eccentric orbit by a passing star, thus becoming able to approach the Oort 
cloud closely at every subsequent perihelions and trigger comet showers.

The median age uncertainty, achieved thus far in the lunar spherule project,
is about 150~Myr not sufficient to resolve a 26~Myr periodicity -- the main
prediction of the Nemesis hypothesis. But future similar studies will 
hopefully reach the necessary precision. If the 26~Myr periodicity in 
cratering rates is unambiguously established but the Nemesis nevertheless
is not found in future parallax surveys of the stars as dim as 10th magnitude
(the Hipparcos satellite surveyed only about 1/4 of the known candidates
(Culler et al. 2000)), the mirror option will get strong support.

Even if mirror solar companions exist and are invisible,
then their existence could still be confirmed!
Even completely dark compact gravitating objects reveal 
themselves through the gravitational lensing effect they produce on 
background stars (Paczy\'{n}ski 1997, Roulet and Mollerach 1997).
It is expected that Space Interferometry Mission (SIM),
planned to be launched in 2005, will allow a determination of the mass, 
the distance, and the proper motion of virtually any MACHO capable of inducing
a microlensing event (Miralda-Escud\'{e} 1996, Paczy\'{n}ski 1998). 
For putative microlensing event due to
Nemesis the angular Einstein ring radius would be (Paczy\'{n}ski 1998)
$$\varphi_E\approx 90~\mathrm{mas}~\sqrt{\frac{M_N}{M_\odot}}~
\sqrt{\frac{1~\mathrm{pc}}{D_N}}\; , $$
\noindent where $M_N$ is the Nemesis mass and $D_N$ the distance to it. Thus
it will be resolved by SIM which is expected to have angular resolution of
about 10~mas. Therefore if such a microlensing event is really detected, it
will give a very detailed information about Nemesis. The only problem is that
because the present position of the Nemesis is unknown we are forced to relay
merely on a chance to discover the event or perform a full sky dedicated
search for it. 

Whether or not mirror matter exists will 
become clearer as time goes by. In the mean time,
it is fun to think about the implications of
fascinating possibilities such as mirror planets in our
solar system. In addition to the (admittedly very speculative)
evidence for faint solar companions provided by 
observations discussed above, it is also possible that
some other much closer and smaller mirror planet
can also exist.  Over time, if its orbit is eccentric enough, such planet 
can approach to various ``normal'' solar planets and cause observed
oddities in the solar system, like Pluto's orbit. We can also speculate
that the formation of the Moon was a result of tidal fission of the 
Earth caused by a close encounter with a mirror planet.

But speculations apart, the hypothesis that there are some mirror 
objects in the solar system is in principle testable hypothesis,
because these mirror objects can lead to observable effects due to
their gravitational interactions and they may also
observably radiate if they contain enough ordinary matter.  


\vskip 0.4cm
\noindent
{\bf Acknowledgement}
\vskip 0.4cm
\noindent
R.F. is an Australian Research Fellow. We also thank
J. B. Murray and J. Matese for correspondence.

\end{document}